\documentclass[12pt]{article}

\pdfoutput=1

\usepackage[makeroom]{cancel}
\usepackage{color}
\usepackage{graphicx}
\usepackage{amsmath}
\usepackage{amssymb}
\usepackage{xspace}
\usepackage[small]{subfigure}
\usepackage[numbers,compress]{natbib}
\usepackage[hyperfootnotes=false]{hyperref}
\usepackage{tcolorbox}

\newlength{\fighskip} \fighskip=2pt
\newlength{\figvskip} \figvskip=3pt

\newcommand*{\figbox}[2]{{
  \def\figscale{#1}
  \def\arraystretch{0.8}
  \arraycolsep=0pt
  \begin{array}{c}
    \vbox{\vskip\figscale\figvskip
      \hbox{\hskip\figscale\fighskip
        \includegraphics[scale=\figscale]{#2}}}
  \end{array}}}

\usepackage{mciteplus} 
\usepackage{dcolumn}
\usepackage{bm}
\usepackage{verbatim}
\usepackage{amscd}
\usepackage{amsfonts}
\usepackage{setspace}
\usepackage{amsthm}
\usepackage{enumerate}
\usepackage{mathtools}

\theoremstyle{plain}

\theoremstyle{plain}

\theoremstyle{plain}
 
\theoremstyle{plain}

\theoremstyle{remark}

\theoremstyle{conjecture}

\theoremstyle{corollary}

\begin{document}

\title{\bf 
Projective measurement of black holes
}
\author{
Beni Yoshida\\ 
{\em \small Perimeter Institute for Theoretical Physics, Waterloo, Ontario N2L 2Y5, Canada} }
\date{}

\maketitle

\begin{abstract}
We study the effect of projective measurements on the entanglement structure of quantum black holes.
It is shown that the entanglement verification in monitored quantum circuits, recently discussed in condensed matter physics, is equivalent to the information recovery from a black hole with projective measurements. This correspondence provides useful predictions about non-perturbative effects on quantum gravity and some insights on the black hole interior as well as the final state proposal.
\end{abstract}



\vspace{-0.7\baselineskip}


\section{Introduction}\label{sec:intro}

It had been commonly believed that projective measurements of local qubits break long-range quantum entanglement. It thus came as a great surprise that monitored quantum circuits (MQCs), consisting of both unitary dynamics and local projective measurements, can retain the volume-law entanglement~\cite{Li:2018aa, Skinner:2019aa, Chan:2019wy}. The conceptual pillar behind MQCs is quantum error-correcting code (QECC) and quantum information scrambling. Indeed, MQCs are dynamical realization of QECCs~\cite{Choi:2020aa, Gullans:2020aa} where initial states are encoded into highly entangled many-body states and local projective measurements cannot destroy the encoded logical qubits easily. Furthermore, long-range entanglement in MQCs emerges dynamically by projective measurements whose effects are amplified due to the operator spreading from scrambling dynamics~\cite{Beni21b}.

The physics of MQCs is motivated by fundamental considerations of many-body physics, which are also likely to suggest a host of further generalizations outside its original context. In this paper, we begin to ask the effects of projective measurements on quantum black holes.

Traditional approaches apply the standard treatment of projective measurements in quantum mechanics on curved spacetime by introducing the Unruh detectors and work within the semiclassical framework. Lessons from MQCs, however, suggest that projective measurements of a few quanta from a black hole can induce drastic changes of the entanglement structure which require non-perturbative quantum gravity considerations beyond semiclassical treatment. The effect of projective measurements in quantum gravity has not been systematically studied except a few works, see~\cite{Numasawa:2016aa, Kourkoulou17, Almheiri2018} for instance. Here we address the problem of projective measurements in quantum gravity by making use of quantum information theory. 

Our main result is to observe that entanglement verification in MQCs are equivalent to the Hayden-Preskill (HP) recovery with projective measurements. With this correspondence in hand, one can translate various results on MQCs into the language of quantum black holes and make some predictions about potential non-perturbative effects. We will also revisit the final state proposal by Horowitz and Maldacena~\cite{Horowitz:2004aa} and show that it can be interpreted as the HP recovery with projective measurements. We, however, do not discuss the geometric nature of non-perturbative effects in this paper.

This paper is organized as follows. In section~\ref{sec:review}, we present a brief review of MQCs. In section~\ref{sec:HP}, we discuss the relation to the HP recovery problem as well as the final state proposal. In section~\ref{sec:outlook}, we provide brief discussions.

\section{Monitored quantum circuits}\label{sec:review}

In this section, we present a review of MQCs. 

\subsection{Quantum error-correcting code}

\begin{figure}
\centering
(a)\includegraphics[width=0.18\textwidth]{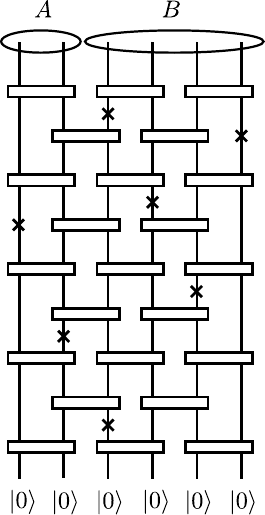} \qquad
(b)\includegraphics[width=0.38\textwidth]{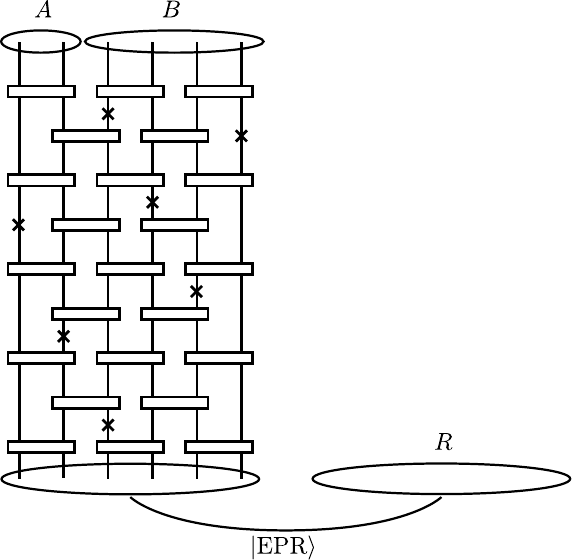}
\caption{MQCs. Blocks represent two-qubit random unitary gates and crosses represent projective measurements. (a)  Pure initial state. (b) Maximally entangled initial state. 
}
\label{fig-review}
\end{figure}

Given a one-dimensional system of $n$ qubits prepared as a product state $|0\rangle^{\otimes n}$, let it evolve by sequential applications of Haar random two-qubit unitary gates on neighboring qubits. We all know that, after the thermalization time of $O(n)$, a highly entangled state with the volume-law entanglement will be created:
\begin{align}
S_{A}\approx n_{A} \qquad \Big(n_{A}<\frac{n}{2}\Big).
\end{align} 
Imagine that, instead, at each step of the aforementioned circuit, a local projective measurement is performed on each qubit with finite probability $p$ (Fig.~\ref{fig-review}(a)). One might guess that the system cannot retain the volume-law entanglement since projections break entanglement associated with measured qubits. Surprisingly, it has been found that, when the measurement rate $p$ is below the threshold $p_{c}$, the system still obeys the volume law~\cite{Li:2018aa, Skinner:2019aa, Chan:2019wy}:
\begin{align}
S_{A}\approx \alpha(p) n_{A}.
\end{align} 

The stability of the volume-law can be understood by interpreting MQCs as QECCs~\cite{Choi:2020aa, Gullans:2020aa}. Let us ask if the initial state can be reconstructed from the output wavefunction. Information theoretically, this question can be addressed by using the Choi state by appending the reference system $R$ and prepare a maximally entangled state (Fig.~\ref{fig-review}(b)). In the present setup of MQCs, the entanglement between the system and the reference can be quantified by the mutual information, corresponding roughly to the number of logical qubits $k$:
\begin{align}
k \approx \frac{1}{2}I(AB,R). 
\end{align}
Strictly speaking, one needs to maximize over all states to find $k$, but we ignore this subtlety.

The number of logical qubits $k$ gradually decreases since projective measurements purify the system. Previous studies, which computed the R\'enyi-$2$ mutual information $I^{(2)}(A,R)$, have found evidences that it will eventually reach some steady value $\approx k_{0}$ after the equilibrium time. It then remains almost constant for an extraordinarily, possibly exponentially, long quantum memory time~\cite{Gullans:2020aa}. In the language of QECCs, this is because the initial conditions are encoded into highly entangled states and local projective measurements cannot destroy them. A codeword subspace of dimension $\sim 2^{k_{0}}$ emerges dynamically from projective measurements and part of initial information can still be recovered from the output. 

The QECC nature of MQCs gives rise to an interesting length scale $d_{\text{code}}$ which can be interpreted as the \emph{code distance} of an MQC~\cite{Li:2021aa}. The physical significance of $d_{\text{code}}$ is profound. When $n_{A}< d_{\text{code}}$, the subsystem $A$ is not large enough to deduce the initial state of an MQC, and thus two subsystems $A$ and $R$ are decoupled~\cite{Hayden:2008aa}:
\begin{align}
\left\Vert \rho_{AR} - \rho_A \otimes \rho_R \right\Vert_{1} \approx 0.
\end{align}
This in turn implies that how two subsystems $A$ and $B$ are entangled is actually independent of the initial condition~\cite{Beni21b}. Hence, the entanglement of an MQC below the $d_{\text{code}}$ scale is \emph{state-independent}. In the literature of MQCs, $d_{\text{code}}$ is often estimated by finding the size of $A$ such that $I^{(2)}(A,R)\approx 0$ instead of $I(A,R)$. This is fine for Clifford dynamics or for qualitative analysis, but may require more careful analysis for generic dynamics.

Given that the output wavefunction still remembers some initial information, the state-independence of the entanglement may sound counterintuitive. Even more surprisingly, the entanglement below $d_{\text{code}}$ is \emph{history-independent} as well, in a sense that it is independent of history of the dynamics which occurred longer than the equilibrium time $t_{\mathrm{equiv}}$ before. This is because an MQC running for $t_{\mathrm{equiv}}$ is sufficient to decouple $A$ from the past. Hence, how two subsystems $A$ and $B$ are entangled with each other is determined only by the dynamics within $t_{\mathrm{equiv}}$ . 

When $n_{A} > d_{\text{code}}$, the subsystem $A$ may be large enough to recover some partial information about the initial states. Hence, the entanglement above $d_{\text{code}}$ depends on the initial states as well as all the history of the dynamics in an MQC, up to an exponentially long memory time. Here, the entanglement structure is state-dependent and history-dependent.

Finally, let us quote some of known results on coding properties of MQCs. For a one-dimensional MQC with random Clifford unitary gates in the volume-law phase, numerical studies have found~\cite{Li:2021aa}:
\begin{align} 
k \sim O(n), \qquad d_{\text{code}} \sim O(n^{\frac{1}{3}}).
\end{align}
So, the emerging QECC has a finite storage rate $r\equiv\frac{k}{n}$ and $d_{\text{code}}$ grows with the system size $n$. For some classes of all-to-all coupled systems, an analytical treatment is possible, which predicts~\cite{Bentsen:2021aa}
\begin{align}
k \sim O(n), \qquad d_{\text{code}} \sim O(n)
\end{align}
in the low $p$ regime. Detailed behaviors of $k$ and $d$ are not essential for our purpose. The most important point is that MQCs can have the volume-law phase with $k\sim O(n)$ and the code distance $d_{\text{code}}$ can be very large. 

\subsection{Simple toy models}

Here we present simple toy models which reproduce some of salient features of MQCs. We begin with the simplest possible toy model of MQCs. Imagine that we initially have a maximally mixed state on $n$ qubits, perform projective measurements on $D$ in the $Z$-basis, and then let the system evolve by a Haar random unitary $U$:
\begin{align}
 \figbox{1.5}{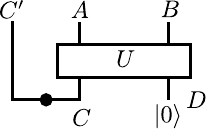}  \label{eq:toy4}
\end{align}
with $C'$ purifying $C$. Here the black dot represents a numerical factor of $\frac{1}{\sqrt{d_{C}}}$ for proper normalization of EPR pairs.

We are interested in whether $A$ and $B$ are entangled or not. A simple Haar calculation suggests
\begin{equation}
\begin{split}
I(A,B)&\approx 2n_{A} \qquad \Big(n_{A} < \frac{n_{D}}{2}\Big).
\end{split}
\end{equation}
Hence, for sufficiently small $A$, two subsystems $A,B$ are nearly maximally entangled. Furthermore, $A$ and $C'$ will be decoupled:
\begin{align}
I(A,C')\approx 0, \qquad \left\Vert \rho_{AC'} - \rho_{A}\otimes \rho_{C'} \right\Vert_{1} 
\leq O\Big( 2^{\frac{2n_{A}- n_{D}}{2}} \Big) 
\end{align}
for $n_{A} < \frac{n_{D}}{2}$ due to the monogamy relation $I(A,C')+I(A,B)=2S_{A}=2n_A$.

A notable consequence of the decoupling phenomena is that the entanglement between $A$ and $B$ will remain unaffected by any operation on $C'$. Namely, if we project $C'$ onto some pure state $|\phi^*\rangle$, then we will have $|\phi\rangle$ on $C$, and $AB$ are still entangled in the same manner, unless $|\phi\rangle$ is fine-tuned and/or $U$ is atypical. Hence, if $U$ is a scrambling unitary and $n_{D}>2n_{A}$, the $AB$ entanglement is independent of the initial states on $C$. In the language of QECCs, this toy MQC has $k= n_{C}$ and $d_{\text{code}}\approx \frac{n_{D}}{2}$.

Next, let the system evolve by Haar random unitary operators $U_{1},\cdots, U_{\tau}$ with projective measurements inserted between them: 
\begin{align}
 \figbox{1.5}{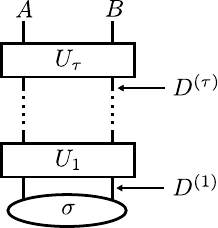} \label{eq:toy2}
\end{align}
where $\sigma$ is an arbitrary initial state, be it pure or mixed. For simplicity of discussion, we assume that $D^{(1)},\cdots,D^{(\tau)}$ consist of the same number of qubits, which is $n_{D}$. 

Let us discuss the cases when $n_{A} < \frac{n_{D}}{2}$. For $\tau=1$, $A$ is decoupled from the reference system and hence the entanglement is state-independent. This implies that, for $\tau>1$, the two subsystems $A$ and $B$ remain maximally entangled regardless of whatever happened before $U_{\tau}$ and the $D^{(\tau)}$ measurements. Indeed, in order for $A$ and $B$ to be maximally entangled, it suffices to have a scrambling unitary $U_{\tau}$ and $n_{D}>2n_{A}$; the detailed properties of $U_{1},\cdots, U_{\tau-1}$, as well as the initial state $\sigma$, are irrelevant. Hence, the entanglement below $d_{\text{code}}\approx \frac{n_{D}}{2}$ is history-independent.

Next, let us consider the case with $n_{A} > \frac{n_{D}}{2}$. For $\tau=1$, $A$ and $B$ are not maximally entangled. Here we ask how long it takes for $A$ and $B$ to be nearly maximally entangled. Haar random calculations suggest
\begin{align}
\tau \approx \exp{\big[\text{const}\cdot(2n_{A}- n_{D})\big]}.
\end{align}
Hence, the mutual information between $A$ and $B$ grows very slowly. Namely, if we take the thermodynamic limit $n\rightarrow \infty$ while holding both $\epsilon_{A} \equiv n_{A}/n$ and $\epsilon_{D} \equiv n_{D}/n$ constant, it will take an exponentially long time to increase the mutual information. Hence, we have the code distance rate $\frac{d_{\text{code}}}{n} = \frac{\epsilon_{D}}{2}$ and the information storage rate $\frac{k}{n} = 1 - \epsilon_{D}$ which remain stable for polynomially long time.

So far, we have considered toy models with Haar random unitary operators. This is a plausible approximation of strongly scrambling dynamics that evolves for longer than the scrambling time. The hallmark of strong scrambling dynamics is the operator growth where any local operator grows to an $O(n)$-body typical operator. A qualitatively similar result can be obtained if the system has been throughly scrambled with decays of OTOCs, see~\cite{Hosur:2015ylk, Yoshida:2017aa} for details. Here we consider MQCs with weaker scrambling dynamics which evolve longer than the local thermalization time, but shorter than the scrambling time. The well-celebrated Lyapunov growth in a black hole occurs only in this time regime.

A weak scrambling dynamics can be modelled by considering an all-to-all coupled system where random two-qubit Haar random gates are applied sequentially on randomly chosen pairs of qubits. Such a random circuit generates scrambling unitary dynamics after $O(\log n)$ time steps. Here we may simply stop the circuit before it gets fully scrambled. In order to further simplify the analysis, we will focus on random Clifford gates and assume that, under the circuit $U$, any single body Pauli operator will grow to $W$-body Pauli operator with randomly distributed supports of qubits. The variance of the size can be of order of the size itself since the size growth is exponential~\cite{Qi:2019aa}, but for our purpose, this crude approximation suffices. We will work on a regime with $1 \ll W \ll n$ and $1 \ll n_{A} < n_{D} \ll n$. A relevant analysis can be found in~\cite{Schuster:aa, Nezami:aa}.

Let us consider an MQC with weak scrambling unitary $U$ in Eq.~\eqref{eq:toy4}. One can compute the mutual information $I(A,B)$ by studying the operator growth. Roughly speaking, a local Pauli operator on $A$ needs to grow large enough to overlap with $D$ and produce a unique pattern of OTOCs~\cite{Yoshida:21a}. We obtain 
\begin{equation}
\begin{split}
I(A,B) &\approx 2n_{A} \qquad \qquad \big( W n_{D} \gtrapprox 2n \big). \label{eq:weak}
\end{split}
\end{equation}
Recall that the initial state of the system can be prepared by measuring qubits on $D$ in the $Z$ basis. A projection operator can be written as $\prod_{j=1}^{n_{D}} \left( \frac{I + Z_{D_{j}}}{2} \right)$ which applies Pauli-$Z$ operators randomly with $\frac{1}{2}$ probability on each qubit. A typical Pauli-$Z$ operator in the projector has weight $\frac{n_{D}}{2}$ which will grow to $\approx \frac{W n_{D}}{2}$-body operator when $\frac{W n_{D}}{2} < O(n)$. Eq.~\eqref{eq:weak} suggests that $A$ and $B$ are nearly maximally entangled when 
\begin{align}
\frac{W n_{D}}{2} \approx n \label{eq:growth}
\end{align} 
where a typical Pauli-$Z$ operator in the projection is fully scrambled. 

The relation in Eq.~\eqref{eq:growth} has an analog in the AdS black hole perturbed by a gravitational shockwave~\cite{shenker2014black}. Writing it as $\frac{1}{n} \frac{n_{D}}{2}W \sim 1$, the LHS corresponds to the horizon shift from the gravitational shockwave. Namely it has the $\frac{1}{n}$ factor, $W = e^{\lambda t}$ as the Lyapunov growth and $n_{D}$ as the entropy carried by the initial perturbation.

\section{Hayden-Preskill recovery}\label{sec:HP}

In this section, we will consider the HP recovery problem with projective measurements.

\subsection{Correspondence}

The HP recovery problem asks whether a piece of quantum information thrown into an old black hole, which is maximally entangled with the early radiation, can be retrieved by having access to both the early and late radiations~\cite{Hayden07}. Information theoretically, this problem can be studied by considering the following wavefunction:
\begin{align}
 \figbox{1.5}{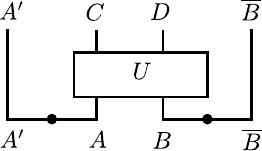}. \
\end{align}
where the old black hole is modelled as $n_{B}$ copies of EPR pairs on $B$ and $\overline{B}$, with $\overline{B}$ being the early radiation. The infalling quantum state is represented by EPR pairs on $A$ and $A'$ where $A'$ is the reference system. The system evolves by some unitary operator $U$, and $C$ and $D$ represent the remaining black hole and the late radiation respectively. The HP recovery problem asks whether quantum entanglement (\emph{e.g.} EPR pairs) can be distilled from $A'$ and $\overline{B}D$.

Here, instead of collecting the late Hawking radiations, let us think of performing projective measurements on them as schematically shown below:
\begin{align}
 \figbox{1.5}{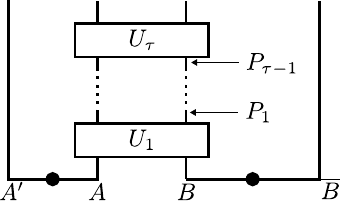} \ .
\end{align}
Then, is the quantum information recoverable from the early radiation $\overline{B}$ ?

To see the correspondence with MQCs, let us turn the diagram upside down~\cite{Beni18}:
\begin{align}
 \figbox{1.5}{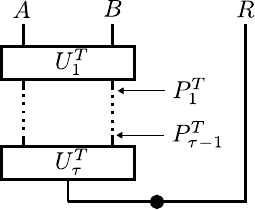} \ 
\end{align}
where we used the following identity:
\begin{align}
\figbox{1.5}{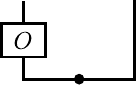} \ = \ \figbox{1.5}{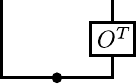}\ ,  \qquad (O \otimes I) |\text{EPR}\rangle = (I \otimes O^{T}) |\text{EPR}\rangle
\end{align}
where $T$ represents the transpose; $O_{ij}\rightarrow O_{ji}$. We then see that the HP recovery with projective measurements is identical to the entanglement distillation in MQCs. Here, the subsystems $A$ and $B$ in MQCs correspond to the infalling quantum state and the early radiation in the HP problem respectively. 

This correspondence can be formally shown. The output wavefunction of an MQC is
\begin{align}
|\Psi(m)\rangle = \frac{1}{\sqrt{\text{Prob}(m)}} \  \figbox{1.5}{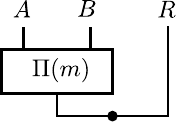} 
\end{align}
where $\Pi(m)$ represents the whole projection circuit with measurement outcomes $m$ and $\text{Prob}(m)$ is the amplitude. We can rewrite this MQC wavefunction as the HP wavefunction:
\begin{align}
|\Psi(m)\rangle = \frac{1}{\sqrt{\text{Prob}(m)}} \ \figbox{1.5}{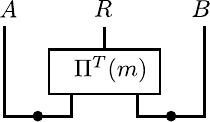} \ .
\end{align}
Hence, the MQC problem and the HP problem are indeed equivalent.

\subsection{Implications from monitored quantum circuits}

Let us look at several cases of the HP recovery problem. Consider the following case:
\begin{align}
\figbox{1.5}{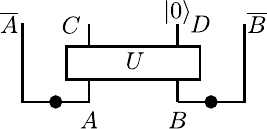} 
\end{align}
where, after the scrambling time, we measure $D$ in arbitrary basis. Let us translate MQC results into the HP setting. When $n_{D}>2n_{A}$, a diary can be reconstructed from the early radiation $\overline{B}$. When $2n_{A}>n_{D}$, a part of the diary will remain inside the black hole $C$. Note that, in the original HP problem, one would need to collect $n_{D}\gtrapprox n_{A}$ qubits. For projective measurements, one needs $n_{D}\gtrapprox 2n_{A}$ qubits. The weak scrambling case offers a similar result where the diary becomes fully recoverable when a typical operator on $D$ grows to a global one with $\frac{1}{n}\frac{n_{D}}{2}W \approx 1$. 

Next, consider the case where projective measurements are performed sequentially:
\begin{align}
 \figbox{1.5}{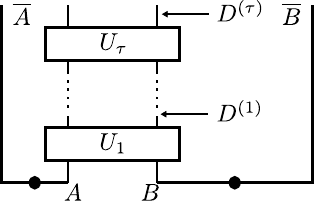}
\end{align}
This corresponds to a ``non-evaporating'' black holes since the total number of qubits $n$ remains constant. This may be achieved by performing measurements of Hawking quanta, and sending them back to the black hole. If the size of diary is smaller than $d_{\mathrm{code}}\approx \frac{n_{D}}{2}$, the diary will become immediately available for reconstruction on $\overline{B}$. If the diary is larger, part of the diary remains inside the black hole for an exponentially long time.

A physical reason why it takes so long to recover a large diary is as follows. When the first projective measurement on $D^{(1)}$ is performed, part of the diary becomes recoverable from $\overline{B}$. But the $D^{(1)}$ measurement creates a new infalling matter. When the second measurement on $D^{(2)}$ is performed, it will reveal the information about the recent input from $D^{(1)}$ instead of the original diary. In general, measurements of $D^{(j+1)}$ tend to make the most recent input $D^{(j)}$ available on $\overline{B}$ while the effect on the original diary decreases exponentially in $\tau$, which leads to a significant delay in recovering $A$. 

Let us now consider a situation close to an actual MQC. Detailed behaviors depend on the specifics of each model, but one can deduce some universal features. Recall that $S_{A}\approx\alpha n_{A}$ with $0 < \alpha <1$. This suggests that a part of the diary will be damaged as a result of projective measurements. For the remaining $\alpha$-portion of the diary, the content below the $d_{\text{code}}$ scale will become immediately available on $\overline{B}$, but the content above $d_{\text{code}}$ stays inside the black hole for exponentially long time. Hence, the recoverability of finite portion in the HP problem indeed stems from the volume-law entanglement in MQCs.

It is, however, incorrect to think that $(1-\alpha)$-portion is permanently lost due to projective measurements since one can prepare ancilla qubits to simulate the measurement process as a unitary event. In this interpretation, the portion of the diary, which appear to be lost, will be encoded into a joint system of ancilla qubits and the black hole. 

That being said, the HP recovery can be actually performed without damaging information by waiting for the scrambling time. An example of such strategies, as well as the conversion into the MQC setting, are depicted below
\begin{align}
\figbox{1.5}{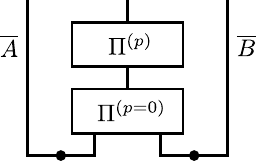} \ \longrightarrow \ \figbox{1.5}{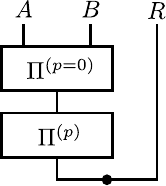}
\end{align}
where $\Pi^{(p=0)}$ represents the circuit with the measurement rate $p=0$. Here, the MQC $\Pi^{(p)}$ purifies the system, keeps roughly $\sim 2^{k_p}$-dimensional code subspace, and then $\Pi^{(p=0)}$ will scramble it. This suggests
\begin{align}
I(A,B) \approx 2n_{A} \qquad \Big( n_{A} \lessapprox \frac{n - k_{p}}{2} \Big),
\end{align}
and the diary will be fully recoverable as long as $n_{A} \lessapprox \frac{n - k_p}{2}$.

HP recovery algorithms can be converted into entanglement verification algorithms in MQCs and vice versa. There are two types of recovery strategies for the HP problem. The algorithm of a first kind postselects the measurement outcome and works probabilistically~\cite{Yoshida:2017aa}. A typical instance of the postselection algorithm, as well as its conversion to the MQC, are depicted below: 
\begin{align}
\figbox{1.5}{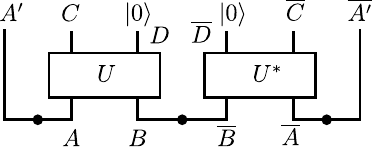} \ \longrightarrow \ \figbox{1.5}{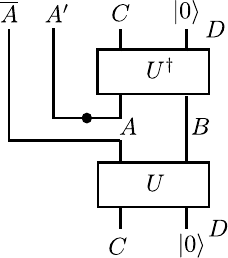}
\end{align}
where measurement outcomes on $D$'s need to match. This algorithm works universally for scrambling systems and can be turned into a deterministic one at the cost of increasing the circuit complexity linearly with $d_{A}$. Entangled states will be distilled on $A'\overline{A'}$ and $\overline{A}A'$ for the HP recovery and MQCs respectively. 

The algorithm of a second kind resembles the traversable wormhole geometry and works in the weak scrambling regime~\cite{Traversable2017, Nezami:aa}. It typically implements a quantum circuit of the following form:
\begin{align}
\figbox{1.5}{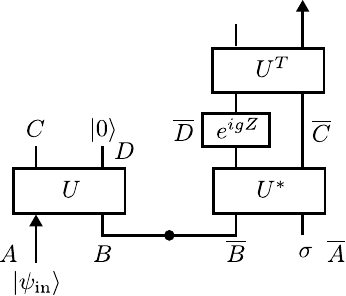} \ \longrightarrow \ \figbox{1.5}{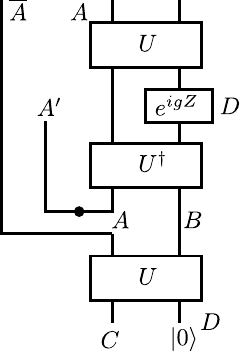}.
\end{align}
where $e^{igZ}$ with some appropriate value of $g$ needs to be applied. Entangled states will be distilled on $\overline{A}A$. A similar conversion was already noted in~\cite{Schuster:aa}. 

\subsection{Final state proposal}

\begin{figure}
\centering
(a)\includegraphics[width=0.24\textwidth]{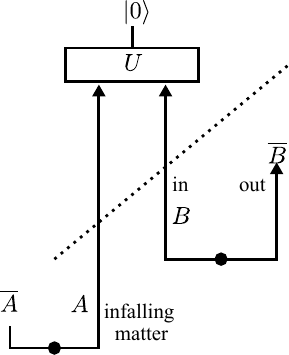} \qquad
(b)\includegraphics[width=0.24\textwidth]{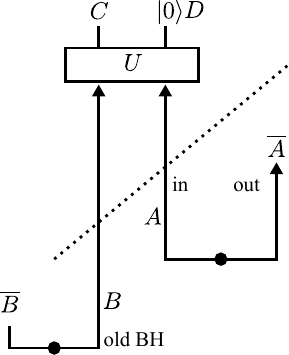}
\caption{Two models of the final state projection. Dotted lines schematically represent the black hole horizon. (a) Unitary dynamics. (b) Page curve behavior.
}
\label{fig-final}
\end{figure}

The final state proposal~\cite{Horowitz:2004aa} attempts to reconcile the Hawking's semiclassical prediction with the unitarity of black holes by postulating that some non-perturbative quantum gravity effect, which is yet to be discovered, postselects the final state of the wavefunction. One possible realization of the proposal, which slightly differs from the original one, can be expressed as in Fig.~\ref{fig-final}(a) where an $n_{A}$-qubit infalling matter on $A$ collapses into a black hole which eventually evaporates by creating $n_{B}$ pairs of entangled Hawking modes. Here a reference system $\overline{A}$ is appended to enable information recovery analysis. The infalling Hawking mode $B$ and the infalling matter $A$ interact with each other and are projected onto some pure state $|0\rangle$.

The overall process can be described approximately by some unitary if $\overline{A}$ is nearly maximally mixed. This is indeed the case if $U$ is strong scrambling unitary and $n_{B}\gtrapprox n_{A}$. Namely, the Page's theorem suggests
\begin{align}
\left\Vert \rho_{A} - \frac{I_{A}}{d_{A}} \right\Vert_{1} \leq O( 2^{ n_{A} - \frac{n}{2} }).
\end{align}
Hence, strong scrambling dynamics, followed by generic projection, can restore the unitarity of the black hole dynamics. 

This version of the final state proposal can be interpreted as the HP recovery where the infalling matter $A$ is recovered on the outgoing modes $\overline{B}$ via projective measurements on the whole system. The only difference is that the projection $|0\rangle\langle 0|$ is somehow forced such that it is consistent with the unitary dynamics seen from the outside. 

Here it is worth making a comment on the issue raised by Gottesman and Preskill~\cite{Gottesman:2004aa}. They pointed out that perturbations inside the black hole, such as interactions between $A$ and $B$ may lead to non-unitary dynamics in the outside. In our formulation of the final state proposal, one does not suffer from this issue since small perturbations cannot undo the scrambling dynamics $U$. 

A modification of this final state model can reproduce the Page curve behavior as well (Fig.~\ref{fig-final}(b)). Here, our interpretation of each subsystem changes as follows. The EPR pairs on $\overline{B}$ and $B$ represents an old black hole with $\overline{B}$ being the early radiation while EPR pairs on $A$ and $\overline{A}$ represents entangled Hawking modes. This time, one needs to apply projections on $n_{D}\approx 2n_{A}$ qubits for a reason discussed below. 

Initially, the black hole is maximally entangled with $S_{\mathrm{ent}} = S_{\mathrm{BH}}=n_{B}$ where $S_{\mathrm{BH}}$ is the coarse-grained (Bekenstein-Hawking) entropy of a black hole. In order for the entanglement entropy $S_{\mathrm{ent}}$ to decrease as a result of emitting the outgoing mode $\overline{A}$, it needs to be entangled with the early radiation $\overline{B}$. Once again, this quantum circuit can be interpreted as the HP recovery where $A$ needs to be recovered on $\overline{B}$. If $U$ is a strong scrambling dynamics, it suffices to take $n_{D}\geq 2n_{A}$ in order for $\overline{A}$ to be entangled with $\overline{B}$. In order to match with the expected Page curve behavior with $S_{\mathrm{BH}} \approx n_{B}-n_{A}$, one actually needs to take $n_{D}\approx 2n_{A}$ such that $S_{C}\approx n_{B}-n_{A}$. If we increase $n_{A}$ to $\approx n_{B}$, we recover the final state model in Fig.~\ref{fig-final}(a). 

\section{Discussions}\label{sec:outlook}

We have presented several predictions about the effect of projective measurements on quantum black holes based on purely quantum mechanical considerations. Measurements of a few Hawking quanta indeed change the entanglement structure drastically. Developing a bulk quantum gravity interpretation, including the non-perturbative effect of projective measurements, is an important future work. Since the projection $|0\rangle \langle 0|$ is a probabilistic application of local Pauli-$Z$ operators, it will create a shockwave which is roughly equivalent to perturbing $\frac{n_{D}}{2}$ qubits. Recalling that the projector $|0\rangle \langle 0|$ is symmetric in time, it is reasonable to think that the shockwave will propagate both to the future as well as to the past, modifying the initial condition such that it is consistent with the measured outcome $|0\rangle$. This way, a diary $A$ will cross the backward shockwave from $D$, realizing a situation close to the geometrization of the OTOC calculation. Perhaps, this line of argument may enable us to relate the proof of information recoverability via OTOCs to another proof based on the entanglement wedge reconstruction. 

We have seen that the entanglement structure in MQCs below the $d_{\text{code}}$ scale is state-independent and history-independent. It is worth noting that entanglement verification problem in MQCs can be interpreted as the interior reconstruction problem in a monitored black hole. Namely, by interpreting the subsystem $A$ as the outgoing Hawking mode, finding the degrees of freedom with which $A$ is entangled is nothing but the entanglement verification. When an old black hole remains unperturbed, the outgoing mode $A$ is entangled with the early radiation $R$ and its interior partner mode is thus state-dependent. When an old black hole is monitored, however, the outgoing mode $A$ will be decoupled from the early radiation $R$, and will be entangled with its complementary subsystem $B$. This feature is particularly appealing since the same construction of an interior partner mode remains valid for an arbitrary initial state, be it pure or mixed. Hence, if a black hole experiences continuous projections under some physical mechanism, state-independent interior operators can be constructed which enables us to avoid a version of the firewall puzzle for black hole typical states due to Marolf and Polchinski~\cite{Marolf:2016aa}. In fact, a certain version of MQCs can be created without performing measurements. One possible mechanism is to consider the effect of introducing an external probe, or an infalling observer, who actually enters the black hole by crossing the horizon~\cite{Beni19, Yoshida:2021aa}. 

We have demonstrated that the flow of quantum information in the final state model resembles the HP recovery with projective measurements. The underlying mechanism behind the final state projection remains unclear. It may be interesting to construct a toy model of evaporating black holes by modelling the thermal decoherence as projections. One possibility is to use MQCs with adiabatically increasing the measurement rate $p$. Such a model may admit state-independent interior modes from projections while, at the same time, exhibiting the Page curve behavior from increasing $p$. 

Other studies on MQCs also suggest possible quantum gravity interpretations. Namely, an effective theory description for computing entanglement entropies, fundamentally akin to the Ryu-Takayanagi formula, has been proposed~\cite{Li:2021aa}. An analytical calculation from~\cite{Bentsen:2021aa} will be useful in developing further insights. Tensor network toy models can provide some insights on the geometrization of measurement effects. In fact, one of the earliest works on MQCs considers such a setup in a hyperbolic space~\cite{Vasseur:2019vm}.

\subsection*{Acknowledgment}

I thank Chris Akers for useful discussions. Research at the Perimeter Institute is supported by the Government of Canada through Innovation, Science and Economic Development Canada and by the Province of Ontario through the Ministry of Economic Development, Job Creation and Trade.

\providecommand{\href}[2]{#2}\begingroup\raggedright\endgroup

%
\end{document}